\pdfoutput=1

\documentclass[11pt, letterpaper]{article}

\newcommand\revision[1]{{#1}}

\usepackage[nodisplayskipstretch]{setspace}
\usepackage[margin={2cm,2cm}]{geometry} 

\usepackage{sectsty}
\usepackage{abstract}
\usepackage{color}
\usepackage{graphicx}
\usepackage{multicol}
\usepackage{amsmath,amssymb,amstext,amsfonts}
\usepackage[usenames,svgnames]{xcolor}
\usepackage{float}
\usepackage{subfig}
\usepackage{caption}
\usepackage{framed}
\usepackage{authblk}
\usepackage{titlesec}
\usepackage{setspace}
\usepackage{newtxtext,newtxmath,fancyhdr}  
\usepackage[normalem]{ulem} 
\usepackage{enumerate} 
\usepackage{mathrsfs} 
\usepackage{bigints}
\usepackage{bm}
\usepackage{enumitem} 
\usepackage{wasysym}
\usepackage{cite}
\usepackage{makecell} 
\usepackage{arydshln} 

\titlespacing*{\section}{0pt}{3ex plus 2ex}{1ex} 
\sectionfont{\fontsize{11}{11}\selectfont} 
\subsectionfont{\fontsize{10}{10}\selectfont} 

\newcommand{\romansubs}{\renewcommand{\theequation}{\theparentequation \roman{equation}}} 

\newcommand{\tinyrmsub}[1]{\mbox{{\tiny{#1}}}}
\newcommand{\scalarfactsym}{1+\epsilon\nabla_{\mu}\phi^{*}g^{(\mu\nu)}\nabla_{\nu}\phi}

\newcommand{\scalarfact}{1+\epsilon\nabla_{\beta}\phi^{*}\nabla^{\beta}\phi}

\newcommand{\VV}{V(\phi^{*}\phi)}
\newcommand{\VVprime}{V^{\prime}\!(\phi^{*}\phi)}
\newcommand{\ourkinterm}{1-\frac{\epsilon\omega^{2} f^{2}}{A}+\frac{\epsilon(\partial_{r}f)^{2}}{B}}

\newcommand{\PRLsep}{\noindent\makebox[\linewidth]{\resizebox{0.750\linewidth}{1pt}{$\blacklozenge$}}\bigskip}
\newcommand{\dd}{\mathrm{d}}
\newcommand{\rhoint}{\rho_{\tinyrmsub{int}}}
\newcommand{\rhoext}{\rho_{\tinyrmsub{ext}}}
\newcommand{\rmax}{r_{\tinyrmsub{max}}}

\usepackage[sort&compress,numbers]{natbib}
\usepackage{xurl}
\usepackage[hypertexnames=false]{hyperref}
\hypersetup{
    colorlinks=true,
    urlcolor=SteelBlue,
    linkcolor=red,
    citecolor=blue,
}

\usepackage{doi}

\begin{document}
\sloppy 
\pagestyle{fancy}
\fancyhead{} 
\fancyhead[OR]{\thepage}
\fancyhead[OC]{{\small{
   \textsf{Born-Infeld boson compact objects}}}}
\fancyfoot{} 
\renewcommand\headrulewidth{0.5pt}
\addtolength{\headheight}{2pt} 
\global\long\def\tdud#1#2#3#4#5{#1_{#2}{}^{#3}{}_{#4}{}^{#5}}
\global\long\def\tudu#1#2#3#4#5{#1^{#2}{}_{#3}{}^{#4}{}_{#5}}

\twocolumn

\title{\vspace{-2cm}\hspace{-0.0cm}\rule{\linewidth}{0.2mm}\\
\bf{\Large{\textsf{An analysis of Born-Infeld boson compact objects}}}}


\author[1,2]{\small{Andrew DeBenedictis}\thanks{\href{mailto:adebened@sfu.ca}{adebened@sfu.ca}}}%

\author[3]{\small{Sa\v{s}a Iliji\'{c}}\thanks{
   \href{mailto:sasa.ilijic@fer.hr}{sasa.ilijic@fer.hr}}}%

%


\affil[1]{\footnotesize{\it{Faculty of Science, Simon Fraser University}}\\
   \footnotesize{\it{Burnaby, British Columbia, V5A 1S6, Canada}} \protect\\
   \footnotesize{and}
}

\affil[2]{\footnotesize{\it{The Pacific Institute for the Mathematical Sciences \vspace{0.2cm}}}
}

\affil[3]{
   \footnotesize{\it{\!University of Zagreb Faculty of Electrical Engineering and Computing,}}\protect\\
   \footnotesize{\it{Department of Applied Physics, Unska 3, 10\,000 Zagreb, Croatia}}
}

\date{\vspace{-0.8cm}({\footnotesize{\today}})} 
\twocolumn[ 
  \begin{@twocolumnfalse}  
  \begin{changemargin}{1.75cm}{1.75cm} 
\maketitle
\end{changemargin}
\vspace{-1.0cm}
\begin{changemargin}{1.5cm}{1.5cm} 
\begin{abstract}
{\noindent\small{In this paper we study gravitationally bound compact objects sourced by a string theory inspired Born-Infeld scalar field. Unlike many of their canonical scalar field counterparts, these ``boson stars'' do not have to extend out to infinity and may generate compact bodies. We analyze in detail both the junction conditions at the surface as well as the boundary conditions at the center which are required in order to have a smooth structure throughout the object and into the exterior vacuum region. These junction conditions, although involved, turn out to be relatively easy to satisfy. Analysis reveals that these compact objects have a richer structure than the canonical boson stars and some of these properties turn out to be physically peculiar: There are several branches of solutions depending on how the junction conditions are realized. Further analysis illustrates that in practice the junction conditions tend to require interior geometries reminiscent of ``bag of gold'' spacetimes, and also hide the star behind an event horizon in its exterior. The surface compactness of such objects, defined here as the ratio $2M/r$, can be made arbitrarily close to unity indicating the absence of a Buchdahl bound. Some comments on the stability of these objects are provided to find possible stable and unstable regimes. However, we argue that even in the possibly stable regime the event horizon in the vacuum region shielding the object is potentially unstable, and would cut off the star from the rest of the universe.}}
\end{abstract}
\noindent{\footnotesize PACS(2010):  04.40.Dg \;\, 04.40.-b \;\, 05.30.Jp}
\rule{\linewidth}{0.2mm}
\end{changemargin}
\end{@twocolumnfalse} 
]
\saythanks 
\vspace{0.5cm}
{\setstretch{0.9} 
\section{Introduction}
The use of scalar fields as a gravitating material source in general relativity has a long history \cite{ref:dasscalars,ref:wymanscalars,ref:santosscalars,ref:mohammadscalars,ref:sotiriouscalars,ref:cloughscalars}. For example, in the realm of cosmology scalar fields have been utilized in order to drive inflation \cite{ref:brandeninflation,ref:eastherinflation,ref:dennisinflation,ref:chervoninflation}, generate an effective cosmological constant \cite{ref:carrollcosconst,ref:suarezcosconst,ref:novellocosconst,ref:khosravicosconst}, and considered as a source for the possible late-time cosmological acceleration \cite{ref:turnerdarkenergy,ref:friemandarkenergy,ref:sergidarkenergy,ref:pacifdarkenergy,ref:marttensdarkenergy}. Stability studies within a cosmological context have been performed in \cite{ref:maxim}. In the arena of high energy particle physics the Higgs field has been postulated, and later realized, to be of scalar nature (see \cite{ref:wuhiggs} and references therein).

Given the possibility of scalar fields in the universe, one may ask if such fields can form condensates due to their self gravitation, much the same way as ordinary matter may form stars. This then gives rise to the interesting field of boson stars. To date, much interesting work has been done in this arena. Much of this work has been done with what we refer to here as canonical scalar fields, sometimes referred to as quintessence in the realm of cosmology. This canonical scalar possesses a standard scalar field kinetic term in the Lagrangian density possibly supplemented with an additive self-interaction term which in the $+2$ metric signature utilized in this manuscript reads as:
\begin{equation}
 \mathcal{L}_{\tinyrmsub{can}}=-\left(\nabla_{\mu}\phi^{*} g^{(\mu\nu)}\nabla_{\nu}\phi + \VV\right)\,. \label{eq:canlag}
\end{equation}
Here $\VV$ the self-interaction potential, assumed to depend only on the product of fields $\phi^{*}\phi$. The explicit symmetrization of the metric is required for the variational principle where variation of the scalar field action with respect to the metric yields its stress-energy tensor, which is the symmetric one in general relativity.

In the majority of the literature involving scalar fields in gravitational studies, it is the canonical scalar field that has been used. Of particular relevance here is the use of scalar fields in the static spherical symmetry, often called ``boson stars''. Such stars have received much attention in the literature. For example, gravitating canonical scalars coupled to non-standard electrodynamic theories have been analyzed in \cite{ref:georgievascalarbh,ref:zhangBIedstar,ref:jaramillobosonstar,ref:brihayebosonstar}. Boson stars in theories with torsion have been studied in \cite{ref:sasachargedtorsionbosonstar}, \cite{ref:sasafoftboson}, and photon spheres have been analyzed in \cite{ref:sasaphotonspherebosonstar}. Non-minimal coupling was also considered in \cite{ref:dubravkobosonstar} and \cite{ref:anjabosonstars}, and some other interesting studies have been performed in \cite{ref:felixselfinteractionbosonstar}. A nice discussion of compact objects in a more general setting may be found in \cite{ref:farookcompactobject}.
Other interesting works have been performed in the phantom regime in \cite{ref:lieblingLRR,ref:dzhuphantomstar,ref:dingphantomstar}.
The field of study involving scalar field gravitational condensates is actually quite vast, prohibiting the citation of all the interesting works, so we refer the interested reader to the thorough reviews \cite{ref:lieblingLRR,ref:jetzerbosonstarreview,Schunck:2003kk} on the subject and references therein.

One interesting property of the massive canonical scalar field is that in static spherical symmetry it does not possess a sharp boundary without the presence of a thin-shell. That is, for these objects to be smooth in the canonical theory, they must fall off to infinity, although the stress-energy can be made arbitrarily small. There is indication though that some exotic potentials can introduce a boundary to the system, thus  possibly creating true compact objects \cite{ref:compactbosons0,ref:compactbosons1,ref:compactbosons2,ref:compactbosons3}.

There is another rather interesting scalar field in the literature which is inspired by Born-Infeld theory and string theory and is known as the Born-Infeld scalar field  \cite{ref:senBIjhep}, \cite{ref:senBImpla}. This field was introduced to describe dynamics of string theory tachyons on D-brane anti-D-brane system or unstable D-branes (see \cite{ref:garoutachy} for early proposals). The Lagrangian density for this type of scalar field is furnished by
\begin{equation}
 \mathcal{L}_{\tinyrmsub{BI}}= -2 \VV\sqrt{\scalarfactsym}\,, \label{eq:tachylag}
 \end{equation}
 where again the explicit symmetrization is used for the same reason as with the canonical scalar field. The peculiar looking factor of $2$ is actually mathematically irrelevant as it simply rescales the potential and also has no overall effect on the resulting scalar field Euler-Lagrange equations of motion, solely yielding an overall factor of 2 in front of them. Its inclusion into the Lagrangian density is purely technical,  so that the variation with respect to $\phi^{*}$ yields exactly the same equations of motion for $\phi$ as the equations derived using the purely real Lagrangian density found in most of the literature, instead of being off by an overall (unimportant) factor of $2$ (assuming that in the explicitly real case the potential depends solely on $\phi^{2}$). In the domain of cosmology, where this scalar field has been studied most extensively from a gravitational perspective, the field is usually taken as real, but there is strictly speaking no need to do so, and hence in (\ref{eq:tachylag}) we generalize it in this work to the complex realm. This allows for a more general and richer structure to the theory. The constant $\epsilon$ in (\ref{eq:tachylag}) can possess the values $+1$ or $-1$, the $+1$ corresponding to the early string theory inspired scalar field, often referred to as the ``tachyon field'' (which in this case has nothing to do with superluminal motion) \cite{ref:sentachy}.

 The field in (\ref{eq:tachylag}) is often called the Born-Infeld scalar field due to the Lagrangian's similarity to that of Born-Infeld electrodynamics \cite{ref:BIorig}. Such scalar fields have been studied extensively in string theory \cite{ref:sentachy} and cosmology
 \cite{ref:luBIcosmology,ref:luphantomBIcosmology,ref:novelloBIcosmology,ref:kernerBIcosmology,ref:kamenshchikBIcosmology,ref:janaBIcosmology}.
 However, surprisingly, unlike the canonical scalar field, almost no work has been done with the Born-Infeld scalar in the vein of gravitational condensates or boson stars save for an interesting constant potential phantom field study in \cite{ref:bilic}. In this manuscript we attempt to modestly fill in this gap somewhat by providing an analytical study of such objects and the properties they must possess within the realm of static spherical symmetry.  

This manuscript is arranged as follows: In section \ref{sec:scalars} we complexify the Born-Infeld scalar field (taken to be real for simplicity in most studies) and construct from this the quantities required for the purposes of the study here. In section \ref{sec:junct} we briefly review the junction condition formalism of general relativity and apply it to the matter-vacuum boundary that a Born-Infeld compact star can possess. We also study the boundary conditions at the center of the star in order for it to have a regular center. This in essence sets up all the conditions required in order to possess a viable solution. We comment on some peculiarities of the solutions. In section \ref{sec:numer} a small numerical model is demonstrated confirming the validity of the work. In section \ref{sec:cosconst} a few comments are made regarding the possibility of including a cosmological constant and finally, in section \ref{sec:conc}, a summary is presented accompanied with some concluding remarks. At the end of the manuscript an appendix is included illustrating some calculations related to the main discussions but which would distract from the main text.

\section{The gravitating complex Born-Infeld scalar field system}\label{sec:scalars}
We consider a self gravitating complex Born-Infeld field minimally coupled to general relativity. The overall action for this physical system is given by
\begin{equation}
 S=\int d^{4}x\, \sqrt{-g}\left[\frac{1}{16\pi}R + \mathcal{L}_{\tinyrmsub{BI}}\right]\,, \label{eq:action}
\end{equation}
$R$ of course being the Ricci curvature scalar. From the above action one can derive all the relevant quantities and equations required for this study, which comprises the Einstein field equations, the stress-energy tensor and the Born-Infeld equation of motion:
\begin{subequations}\romansubs
\begin{align}
 G^{\mu}_{{\;\;}\nu} &= 8\pi T^{\mu}_{{\;\;}\nu}\,,\label{eq:einsteq} \\
 T^{\mu}_{{\;\;}\nu} &= 2\VV \left[\frac{\epsilon\nabla_{(\alpha}\phi^{*}\nabla_{\nu)}\phi\,g^{\alpha\mu}}{\sqrt{\scalarfact}} \right. \nonumber \\
 &\qquad\qquad \left.- \sqrt{\scalarfact}\,\delta^{\mu}_{\;\;\nu}\right]\,, \label{eq:set} \\
 0 &=\nabla_{\sigma}\left[-\frac{\epsilon\VV}{\sqrt{\scalarfact}}\partial^{\sigma}\right] \phi \nonumber \\
 & \qquad \;\; -2\sqrt{\scalarfact}\, \VVprime \phi\,, \label{eq:BIeqn}
\end{align}
\end{subequations}
where $\VVprime:=\partial \VV/\partial(\phi^{*}\phi)$ so that
\begin{equation}
 \frac{\partial\VV}{\partial \phi^{*}}= \VVprime \phi \,, \nonumber
\end{equation}
which has been used in order to derive (\ref{eq:BIeqn}).

In this manuscript we will restrict ourselves to static spherical symmetry mainly using a coordinate chart where the metric takes on the standard form given by the line element
\begin{equation}
 ds^{2}=-A(r)\,{\dd}t^{2}+B(r)\,{\dd}r^{2} +r^{2}\,{\dd}\theta^{2} + r^{2}\sin^{2}\theta\, {\dd}\varphi^{2}\,. \label{eq:lineelement}
\end{equation}
The metric function $B(r)$ is often expressed
in terms of the often-called mass function $m(r)$ defined by
\begin{equation}
B(r) = \frac{1}{1-\frac{2m(r)}{r}} \, .
\end{equation}
The mass function has the property that, for compact objects,
its value at the surface of the object coincides with the Schwarzschild mass.
The ratio $2m(r)/r$ is known as the local compactness,
while its value at the surface of a compact object
is known as the surface compactness of the object. It turns out that unlike with the canonical scalar field, the Born-Infeld scalar field can seemingly form compact bodies for any reasonable potential.

Although the original motivations for the Born-Infeld scalar were time dependent (only), if such a field truly exists in the universe there is no a priori reason that it cannot gravitationally collapse to form a static condensate.

Allowing for a complex Born-Infeld scalar allows us to utilize a time harmonic ansatz for the field,  as is often used in the corresponding theory with canonical scalars, that is
\begin{equation}
 \phi=e^{-i\omega t} f(r) \,. \label{eq:scalaransatz}
\end{equation}
with $f(r)$ real. This restriction on the field is universally used in the study of canonical boson stars so we do not relax it here in order to compare results here with their canonical counterparts in the literature.
Although the field is complex, the ansatz is sufficient to yield a real stress-energy tensor in the Born-Infeld scenario as it also is with the canonical scalar field.

Using (\ref{eq:lineelement}) and (\ref{eq:scalaransatz}) in (\ref{eq:einsteq} - \ref{eq:BIeqn}) yields the following expressions:
\begin{subequations}\romansubs
\allowdisplaybreaks{\begin{align}
-T^{t}_{\;\;t} & =\varrho = 2\VV \frac{B+\epsilon(\partial_{r}f)^{2}}{B\sqrt{\ourkinterm}}\,, \label{eq:rho} \\
T^{r}_{\;\;r} & = p_{r}= -2\VV \frac{A-\epsilon\omega^{2}f^{2}}{A\sqrt{\ourkinterm}}\,, \label{eq:pr} \\
T^{\theta}_{\;\;\theta} &= p_{\perp} = -2\VV\sqrt{\ourkinterm}\,, \label{eq:pt}
\end{align}}
\end{subequations}
The quantities $\varrho$, $p_{r}$ and $p_{\perp}$ are the energy density, radial pressure and transverse pressure respectively as measured by a static observer. We note that if the field is time dependent only ($\partial_{r}f(r)=0$) then the field appears isotropic to the static observer ($p_{r}=p_{\perp}$). In more general scenarios with radial dependence the system appears anisotropic to the static observer, meaning $p_{r}\neq p_{\perp}$.

Although it may not be obvious, the conservation law,
\begin{equation}
\nabla_{\mu}T^{\mu}_{\;\;\nu}=0\,, \label{eq:conslaw}
\end{equation}
which of course is guaranteed once Einstein's equations hold, actually implies that the Born-Infeld equation of motion (\ref{eq:BIeqn}) is satisfied. Therefore in this system of equations it is in principle possible to only solve Einstein's equations in order to solve all equations.
On the other hand, the Born-Infeld equation of motion will imply that the conservation law holds but only where $\partial_{r}f(r)\neq 0$.

Since we have not introduced a gauge field into the  system the action formed by Lagrangian density (\ref{eq:tachylag}) is only globally invariant under a U(1) transformation $\phi\to e^{-i\varepsilon}\phi$. The current associated with this symmetry is given by
\begin{equation}
 J^{\mu}=-\frac{2i\epsilon\VV}{\sqrt{\scalarfact}}\left[\phi^{*}\partial^{\mu}\phi -\partial^{\mu}\phi^{*}\,\phi\right]\,, \label{eq:BIcurrent}
\end{equation}
which for field ansatz (\ref{eq:scalaransatz})
corresponding to a static system,
only has non-zero $0^{\mathrm{th}}$ component.
Its conservation in our static scenario simply implies
time independence of the current density.


\section{Central boundary conditions and surface junction conditions} \label{sec:junct}
We study here the physical properties that the boson stars must possess in order to meet acceptable boundary and junction conditions. To summarize, these conditions will be the following: One condition is that we will require regularity at the center of the star, meaning that the orthonormal components of the Riemann curvature tensor do not blow up there. (It turns out this will enforce other regularity conditions at the origin as well, as we will briefly discuss below.) The other condition is that, at the stellar boundary, the object smoothly joins to the exterior Schwarzschild vacuum spacetime, smoothly here meaning that there is no discontinuous delta-function shell of matter at the surface of the object, and also that there are no pathologies in the Born-Infeld equation of motion there.

Within the realm of general relativity there are several well known and well studied junction conditions. Two of the most important are the Israel-Sen-Lanczos-Darmois (ISLD) junction conditions \cite{ref:I,ref:S,ref:L,ref:D} and the Synge junction conditions \cite{ref:syngeobrien}. The ISLD conditions demand that at a junction in spacetime the induced metric on the surface, $\tilde{g}_{ij}$, and the extrinsic curvature across the junction surface, $K_{ij}$, are continuous as one approaches the junction from one side or the other, i.e.
\begin{equation}
[\tilde{g}_{ij}]_{\pm}=0\, \quad \mbox{and} \quad  [K_{ij}]_{\pm}=0\,, \label{eq:ISLD}
\end{equation}
where $[..]_{\pm}:=(..)_{r^{*}_{+}} - (..)_{r^{*}_{-}}$.
Further, the induced metric on the junction surface is furnished by
\begin{equation}
\tilde{g}_{ij} = g_{\mu\nu}\,h_{i}^{\;\mu}h_{j}^{\;\nu}\,, \label{eq:indmet1}
\end{equation}
where $h_{i}^{\;\mu}:=\partial x^{\mu}/ \partial \tilde{x}^{i}$ with  $\tilde{x}^{i}$ the coordinates on the junction surface. In the case considered here, for an $r=r_{*}=\mbox{const.}$ surface, the calculation simply yields:
\begin{equation}
 \tilde{g}_{tt}=-A(r_{*})\,, \;\; \tilde{g}_{\theta\theta}=r_{*}^{2} \,, \;\; \tilde{g}_{\phi\phi}=r_{*}^{2}\sin^{2}\theta\,. \label{eq:indmet2}
\end{equation}
The extrinsic curvature components are given by
\begin{equation}
 K_{ij}=\nabla_{\mu}\hat{n}_{\nu}\,h_{i}^{\;\mu}h_{j}^{\;\nu}\,, \label{eq:extcurve}
\end{equation}
where $\hat{n}_{\mu}$ is a unit (co)normal vector to the junction surface, $\hat{n}_{\mu}=\sqrt{B}\,\delta^{r}_{\;\mu}$. For the constant $r=r_{*}$ surface of interest here, the non-zero extrinsic curvature components are furnished as
\allowdisplaybreaks{\begin{align}
 K_{tt} & =-\frac{\partial_{r}A(r_{*})}{2\sqrt{B(r_{*})}}\,, \;\; K_{\theta\theta} =\frac{r_{*}}{\sqrt{B(r_{*})}}\,, \nonumber \\ K_{\phi\phi} & =\frac{r_{*}}{\sqrt{B(r_{*})}}\sin^{2}\theta\,. \label{eq:ourextcurv}
\end{align}}
The conditions (\ref{eq:ISLD}) are sufficient to guarantee that there is no infinitely thin shell of matter residing on the junction surface \cite{ref:poissonbook}.

The other well known junction condition of general relativity is that of Synge and O'Brien \cite{ref:syngeobrien}. This junction condition states that, aside from continuity of the metric,
\begin{equation}
 [T^{\mu}_{\;\;\nu}\hat{n}^{\nu}]_{\pm}=0\,. \label{eq:synge}
\end{equation}
It turns out that in the scenario of static spherical symmetry on an $r=\mbox{const.}$ surface, the ISLD junction conditions imply the Synge conditions \cite{ref:DDTpaper}. However, without loss of generality it will be useful in much of the analysis that follows to look at the Synge condition independently of the ISLD conditions.

Before proceeding to the specifics at the boundaries of the object, let us look at the boundary terms that arise from the action (\ref{eq:action}) specifically when we vary the action with respect to $\phi^{*}$ in order to get the Born-Infeld equation of motion for $\phi$. This variation yields the bulk terms, which imply the equation of motion (\ref{eq:BIeqn}) plus a boundary term (B.T.) on the domain boundary, $\partial D$, which reads as
\begin{align}
 \mbox{B.T.}&=-\int_{\partial D} d^{3}\Sigma_{\mu}\, \epsilon\VV \left[\scalarfact\right]^{-1/2} \nonumber \\
 & \qquad\qquad \times g^{(\mu\nu)}\nabla_{\nu}\phi\, \delta\phi^{*}\,. \label{eq:EOMboundaryterm}
\end{align}
Here
\begin{equation}
 d^{3}\Sigma_{\mu}=\sqrt{|g^{(3)}_{\partial D}|}\, d^{3}x\, \hat{n}_{\mu} \,,
\end{equation}
with $g^{(3)}_{\partial D}$ the metric on the boundary. The reason that (\ref{eq:EOMboundaryterm}) is of interest here is that at any terminal boundaries, meaning boundaries that one does not go beyond, which could be infinity, this term should vanish. On the other hand, at any non-terminal boundaries, such as the matter-vacuum junction, $r=r_{*}$, of the compact object, this term should be continuous as one approaches the boundary from one side or the other. (One could view the the interior of the star as the entire domain and not consider the exterior vacuum, in which case the junction surface becomes a terminal boundary, and in that case (\ref{eq:EOMboundaryterm}) should vanish there.) We shall refer to boundary conditions derived from demanding the continuity/vanishing of (\ref{eq:EOMboundaryterm}) as variationally admissible junction conditions. See \cite{ref:jesse} for a more thorough explanation of the above continuity requirement. Most commonly in physics when this issue is faced is that the field value is specified at the boundary so that $\delta\phi^{*}_{\partial D}=0$ (Dirichlet problem) or that the normal derivative at the boundary vanish so that $\hat{n}_{\mu}g^{\mu\nu}\nabla_{\nu}\phi$ vanish there (a specific Neumann boundary condition), both of which will satisfy the vanishing of (\ref{eq:EOMboundaryterm}). Continuity of $\phi$ and/or continuity of its radial derivative are the junction conditions that will be employed to satisfy these.

We will now discuss various boundary and junction conditions both in the gravitational sector and the Born-Infeld sector of the theory for the specific scenarios of interest for this manuscript.

\subsection{Conditions at the center}\label{subsec:center}

One of the main issues we are interested in with regards to the spherical center is the elimination of any curvature singularities. In order to study this aspect of the star we will use the Riemann tensor components projected into an orthonormal frame. We pick the following orthonormal tetrad, $e^{\hat{\alpha}}_{\;\;\mu}$, in order to facilitate this:
\begin{align}
 e^{\hat{t}}_{\;\;\mu} & =\delta^{t}_{\;\mu}\, \sqrt{A},\,\, e^{\hat{r}}_{\;\mu}=\delta^{r}_{\;\mu}\, \sqrt{B} \nonumber \\
  e^{\hat{\theta}}_{\;\mu} & =\delta^{\theta}_{\;\mu}\, r,\,\, e^{\hat{\phi}}_{\;\mu}=\delta^{\phi}_{\;\mu}\, r\sin\theta\,. \label{eq:orthtet}
\end{align}
This leads to the following non-zero orthonormal Riemann tensor components, $R_{\hat{\alpha}\hat{\beta}\hat{\gamma}\hat{\delta}}$, plus those related by (anti)symmetry:
\begin{align}
R_{\hat{t}\hat{r}\hat{t}\hat{r}}
   & = - \frac{A\partial_{r}A\partial_{r}B+B\left((\partial_{r}A)^{2}
     - 2A\partial_{r}^{2}A\right)}{4A^{2}B^{2}}\,, \label{eq:orthoriem} \\
R_{\hat{t}\hat{\theta}\hat{t}\hat{\theta}}
   & = \frac{\partial_{r}A}{2rAB}, \quad
R_{\hat{r}\hat{\theta}\hat{r}\hat{\theta}}
   = \frac{\partial_{r}B}{2rB^{2}}, \quad
R_{\hat{\theta}\hat{\phi}\hat{\theta}\hat{\phi}}
   = \frac{B-1}{r^{2}B}\,.  \nonumber
\end{align}
From the above components $R_{\hat{t}\hat{\theta}\hat{t}\hat{\theta}}$
and $R_{\hat{r}\hat{\theta}\hat{r}\hat{\theta}}$,
we can immediately see that for finite metric functions $A$ and $B$ at $r=0$ we must require that $\partial_{r}A$ and $\partial_{r}B$ vanish at the center in order to avoid infinite curvature there.

Next we note the Tolman-Oppenheimer-Volkoff equation for anisotropic spherical systems:
\begin{align}
 \partial_{r}p_{r}(r)&=-\frac{\varrho(r)-p_{r}(r)}{2r}\left[1-B(r)\left(1+8\pi r^{2}p_{r}(r)\right)\right] \nonumber \\
 &\quad -\frac{2\left[p_{r}(r)-p_{\perp}(r)\right]}{r}\,. \label{eq:anisoTOV}
\end{align}
The metric function $B(r)=1$ at the origin from demanding finiteness of the orthonormal Riemann components (\ref{eq:orthoriem}). Therefore, the only solution to the above equation at $r=0$ which does not give infinite $\partial_{r}p(r)=0$ at the center is one where $p_{r}(r)-p_{\perp}(r)=0$.  In other words, the system must be isotropic at the center. Using this fact we set (\ref{eq:pr}) equal to (\ref{eq:pt}) at $r=0$ and note from (\ref{eq:pr}) and (\ref{eq:pt}) that the quantity $p_{r}(r)-p_{\perp}(r)$ is furnished by
\begin{equation}
p_{r}(r)-p_{\perp}(r)=-2\epsilon \frac{\VV(\partial_{r}f(r))^{2}}{B(r)\left[\ourkinterm\right]}\,. \label{eq:aniso}
\end{equation}
We see that for the above anisotropy to vanish at $r=0$ either $\partial_{r}f(r)=0$ or $\VV=0$ there. $\VV=0$ at the center yields zero central energy density via (\ref{eq:rho}) so it is arguably less physical than imposing the $\partial_{r}f(r)=0$ condition.
In summary, we have shown that for regularity at the center we require that $\partial_{r}A$, $\partial_{r}B$ and $\partial_{r}f$ (or $\VV$) all vanish there. Also, the metric functions $B(r)=1$ and $A(r)>0$ at the center. Further, from spherical symmetry it will be assumed that the geometry is described by even functions in the neighborhood of the center.

\subsection{Conditions at the outer surface}\label{subsec:surface}

At the matter-vacuum junction, $r=r_{*}$, we will first study the ISLD conditions (\ref{eq:ISLD}).
The ISLD junction conditions for the case here boil down to the following continuity conditions at the surface:
\allowdisplaybreaks{\begin{align}
 [A(r)]_{\pm} & = 0\,, \quad [B(r)]_{\pm}=0\,, \nonumber \\
 [r]_{\pm} & = 0, \quad [\partial_{r}A(r)]_{\pm}=0\,, \label{eq:ourISLD}
\end{align}}
The continuity of $r$ in the above may seem obvious but we should note that in more general coordinate systems the ISLD conditions do \emph{not} necessarily demand that the inner and outer coordinate values coincide for a smooth patching \cite{ref:poissonbook}. This will be elaborated on in section \ref{sec:numer}.

Since the spherically symmetric vacuum that we are patching to must be Schwarzschild, the above junction conditions enforce that
\allowdisplaybreaks{\begin{align}
 A(r_{*})& =1-\frac{2M}{r_{*}}\,, \quad B(r_{*})=\frac{1}{1-\frac{2M}{r_{*}}}\, \nonumber \\
 \partial_{r}A(r)_{r=r_{*}} & =\frac{2M}{r^{2}_{*}}\,, \label{eq:schwjunct}
\end{align}}
where $r_{*}$ is the junction location. We have also tacitly assumed that coordinate time corresponds to the proper time of a clock at infinity. The quantity $M$ of course represents the total mass of the star.

The Synge condition (\ref{eq:synge}), which as mentioned previously is satisfied in the scenario here if conditions (\ref{eq:ourISLD}) are met, requires the following on an $r=r_{*}$ surface:
\begin{equation}
 [T^{r}_{\;\;r}]_{\pm}  = 0\,, \nonumber
\end{equation}
which, since the exterior spacetime is vacuum, implies via (\ref{eq:pr}) that $p_{r}(r_{*})$ vanishes at the surface yielding the following subsidiary requirements
\begin{equation}
 p_{r}(r_{*})=0\, \implies \, \left\{\begin{array}{l}
                                                     \epsilon\omega^{2}f^{2}(r_{*}) = \left(1-\frac{2M}{r_{*}}\right)\\[0.2cm]
                                                    \mbox{or,} \\[0.2cm]
                                                    \VV_{|r_{*}}=0\,.
                                                   \end{array}\right. \, \label{eq:fsynge}
\end{equation}
The second condition above makes all components of the stress-energy tensor (\ref{eq:set}) vanish at the boundary, but it will be shown next that the potential can be non-zero inside the star at the boundary yet not violate the junction conditions which patch to an exterior vacuum. In such a case the first condition would be required for smooth patching to vacuum. For the boson star harmonic ansatz of (\ref{eq:scalaransatz}) the first scenario would rule out the $\epsilon=-1$ sector since $r$ should be greater than $2M$. Also, the first scenario tells us that for a purely real scalar field ($\omega=0$) any patching to the vacuum would occur on an event horizon (i.e. $r=2M$ for the vacuum), and no compact Born-Infeld boson ``star'' in the traditional sense could exist unless $\VV=0$ on the surface, which corresponds to the second condition above.

Here the suitability of the first condition in (\ref{eq:fsynge}) will be addressed. We point out that in section~\ref{sec:numer} no solutions were found that obey the second condition, so this first condition may very well turn out to be the more important of the two. It is easy to get lost in the analysis which follows so we will summarize first what is going to be achieved. We will show that in the vacuum domain the only acceptable solution is $\VV=0$. It will then be shown, however, that the potential inside the star at the boundary does not need to vanish, and therefore a patching to Schwarzschild is also possible for Born-Infeld stars with non-zero potential at their surface. In other words, both conditions in (\ref{eq:fsynge}) are acceptable at the boundary, where in the first case the potential from the inside does not need to vanish and is allowed a jump-discontinuity. Such jump discontinuities are not unusual as it is common to have stellar solutions in general relativity where the radial pressure vanishes at the stellar surface but the other components of the stress-tensor posses discontinuities there \cite{ref:thornedisc,ref:saibaldisc,ref:dasdebbook}.

Let us first study the Born-Infeld equation of motion (\ref{eq:BIeqn})
in the Schwarzschild vacuum domain.
The vacuum domain is usually defined as $r>r_{*}$,
whereas the non-vacuum stellar region
is normally defined as $0\leq r \leq r_{*}$.
However, we will find below that Born--Infeld stars
generally tend to possess a peculiar global geometry
and this simple definition is not always valid.
The boundary analysis performed in this manuscript will be local though and so will be insensitive to this peculiarity.

Since the first condition in (\ref{eq:fsynge}) required that $\epsilon=1$ we will be assuming it here. In the Schwarzschild spacetime we want the stress-energy tensor (\ref{eq:set}) to vanish. One obvious way is to set $\VV=0$, but we wish to find out if this is truly the only possibility. It may seem obvious that there is no other option, since looking at (\ref{eq:rho}-\ref{eq:pt}) seems to indicate that if $p_{\perp}$ goes to zero, the other two components become infinite  unless $\VV=0$, due to their denominators going to zero. However we will be a bit careful here in the analysis, since the numerators of these other two components could vanish, negating the possible pathology. 
Now, for $\VV\neq 0$, in order for the numerators of $\varrho$ (\ref{eq:rho}) and $p_{r}$ (\ref{eq:pr}) to vanish, recalling that junction conditions (\ref{eq:ISLD}) require that $A$ and $B$ be set to their Schwarzschild values (and that for $\VV\neq 0$ we must be in the $\epsilon=+1$ branch from (\ref{eq:fsynge})), we must have that
\begin{subequations}\romansubs
\begin{align}
\left(\partial_{r}f(r)\right)^{2}=-\frac{1}{1-\frac{2M}{r}}\,, \label{eq:schwTtt} \\
\omega^{2}f^{2}(r)=1-\frac{2M}{r} \,,\label{eq:schwTrr}
\end{align}
\end{subequations}
and we see that (\ref{eq:schwTtt}) has no solution for the ansatz (\ref{eq:scalaransatz}) which has real $f(r)$. We therefore conclude that in the vacuum domain, the only reasonable condition is $\VV=0$. We will next show that this does not imply that $\VV$ at the stellar boundary needs to be zero. For this to be allowed we will have a jump-discontinuity in $\VV$ as one crosses the boundary from the matter to the vacuum regime, in order for $\VV$ to go from a non-zero value at the surface, to a zero value in the vacuum. This is permitted as long as the pressure vanishes at the boundary without discontinuity, which (\ref{eq:fsynge}) enforces. This then means that in $\VV$, seen as a function of $r$ through its dependence on $\phi(r)$, there is a distributional derivative proportional to a delta ``function'' at the boundary. We will loosely refer to this as an ``infinity in $\partial\VV/\partial r$'' there. Regardless of the terminology, it is a pathology which we do not want in the equations unless they are permitted by acceptable junction conditions. The continuities demanded by the ISLD and Synge junction conditions are designed to take care of the gravitational sector's allowable discontinuities, so we will now concentrate on the Born-Infeld equation of motion.

To analyze the effects of a formally infinite derivative of the potential at the boundary, let us look at the derivative of the potential with respect to radius
\begin{align}
 \frac{\partial\VV}{\partial r} & =\frac{\partial \VV}{\partial (\phi^{*}\phi)} \left(\frac{\partial \phi^{*}}{\partial r} \phi + \phi^{*}\frac{\partial \phi}{\partial r}\right) \nonumber \\
 & = 2 \VVprime\, \partial_{r}f(r)\,. \label{eq:dVdr}
\end{align}
For the above to be infinite at the boundary either $\partial \VV/\partial (\phi^{*}\phi)$ becomes infinite or $\partial_{r}f(r)$ becomes infinite there (or both). Let us now consider the Born-Infeld equation of motion (\ref{eq:BIeqn}) in the non-vacuum region, but at the boundary. This equation, subject to (\ref{eq:scalaransatz}), and then evaluated at the boundary by setting $f(r_{*})$ to condition 1 of (\ref{eq:fsynge}) and $A$ and $B$ to their Schwarzschild values as dictated by (\ref{eq:schwjunct}), yields the following
\begin{align}
&\biggl\{e^{-i\omega t}\partial_{r}f(r_{*})\left[Mr_{*} +\left(\partial_{r}f(r_{*})\right)^{2}\left(6M^{2}-7Mr_{*}+2r_{*}^{2}\right) \right. \nonumber \\
& \left. -2\,\partial_{r}f(r_{*})\,r_{*}^{5/2}\sqrt{r_{*}-2M}\,\omega\right]\VV\biggr\} \nonumber \\
& \div \biggl\{2r^{3}_{*}\left[\left(\partial_{r}f(r_{*})\right)^{2} \frac{r_{*}-2M}{r_{*}}\right]^{3/2}\biggl\} =0. \label{eq:BIboundary}
\end{align}
Notice that in the Born-Infeld equation of motion at the boundary, (\ref{eq:BIboundary}), the problematic factor $\partial_{r}f(r_{*})$ appears in several places. \emph{If} this term is infinite, there is still a chance that the equation may hold since this factor's power in the denominator is 3 and highest power in the numerator is also 3. The equation can therefore still possibly be made to vanish in this case provided that the stellar radius has one of the two values
\begin{equation}
 r_{*}=2M\, \quad \mbox{or} \quad r_{*}=\frac{3}{2}M\,, \nonumber
\end{equation}
both of which would be unacceptable since $r$ must be greater than $2M$. An infinite derivative of $f(r)$ must therefore be avoided at the boundary, and discontinuities in $\partial_{r}f(r)$ are also problematic there. This leaves as the only alternative an infinite value of $\partial \VV/\partial(\phi^{*}\phi)$ across the junction if $\VV_{-}\neq 0$ there (or more precisely a jump-discontinuity in the potential there). This can easily be achieved. For a specific example, the potential could be of the form $\VV=C_{0}\phi^{*}\phi = C_{0}f^{2}(r)$ with $C_{0}=1$ on one side of the junction and $C_{0}=0$ on the other side. The self-interaction switches off in the vacuum although the field itself may remain non-trivial with no measurable consequences. This seemingly strange phenomenon of non-vanishing field with zero stress-energy is somewhat reminiscent of electrodynamics where non-zero $A^{\mu}$ can give rise to zero electric and magnetic fields, and hence no stress-energy.

The jump-discontinuity in $\VV$ but with continuous $\phi^{*}\phi$ is also permitted by the continuity/vanishing of (\ref{eq:BIboundary}) since, for example, by applying the Dirichlet conditions at the junction will make the boundary term vanish at the junction (both approaching from the inside and the outside, but outside the boundary term is zero anyway due to the vanishing of the potential there).

One other case which should be avoided at the surface is that where $\partial_{r}f(r)$ becomes zero. Not only is (\ref{eq:BIboundary}) potentially problematic there, due to the terms which have a higher power of  $\partial_{r}f(r)$ in the denominator than in the numerator, but also the energy density at the surface will diverge if $\VV$ is not zero at the boundary. The energy density (\ref{eq:rho}) at the surface, when junction condition 1 of (\ref{eq:fsynge}) is utilized yields
\begin{equation}
 \varrho(r_{*})=\frac{2\VV}{\left|\partial_{r}f(r_{*})\right|\sqrt{1-\frac{2M}{r}}} \,, \label{eq:rhobound}
\end{equation}
which will diverge when $\partial_{r}f(r_{*})=0$ if $\VV\neq 0$ on the boundary.

To summarize, we have the following conditions that must hold at the junction surface: $A(r)$, $B(r)$ and $\partial_{r}A(r)$ need to be continuous as one crosses the junction. Further, if $\VV\neq0$ as one approaches the junction from the inside we must have that  $f(r_{*})$ is given by the first condition in (\ref{eq:fsynge}) and only $\epsilon=1$ is allowed using ansatz (\ref{eq:scalaransatz}). In order for the Born-Infeld equation to be defined when $\VV\neq 0$ at the junction,  $\partial\VV/\partial(\phi^{*}\phi)$ must be formally infinite there with $\phi$ continuous, yielding non-infinite $\partial_{r}f(r)$ at the surface. Further, $\partial_{r}f(r)$ at the junction cannot equal zero.

If $\VV=0$ at the junction then both $\epsilon=1$ and $\epsilon=-1$ are allowed. It is of interest to note though that, as illustrated in the next section, no solutions were generated that met this condition. It seems that the condition $p_{r}=0$ is more easily achieved via the first condition in (\ref{eq:fsynge}) and not the vanishing of the potential.

\revision{
We summarize all possible boundary and junction conditions in table \ref{table:schwbcs}.

\begin{table}[htbp]
\begin{center}
\begin{tabular}{ |p{3cm}||p{3.5cm}|  }
 \hline
 \multicolumn{2}{|c|}{\makecell{Boundary and junction conditions \\ -spherical chart-}} \\
 \hline
 Location & Requirement \\
 \hline
 {\vspace{0.4cm}center ($r=0$)}   & ${\partial_{r}f(r)=0}$ ${(\mbox{or } \VV=0)}$,\, ${A(r)\neq 0}$,\qquad\, ${\partial_{r}A(r)=0}$, \, ${B(r)\neq 0}$, \,  ${\partial_{r}B(r)=0}$. \\[0.4cm] \hdashline
 {\vspace{1.0cm}matter-vacuum boundary ($r=r_{*}$)} & {\vspace{0.0cm}${[r]_{\pm}=0}$, \qquad\, ${\partial_{r}f(r)\neq 0}$ ${\mbox{if } \VV\neq 0}$, \, ${\partial_{r}f(r)\not\to\infty}$ ${(\mbox{if } \VV\neq 0)}$,\,
 ${[A(r)]_{\pm}=0}$, \qquad \, ${[\partial_{r}A(r)]_{\pm}=0}$, \, ${[B(r)]_{\pm}=0}$,  \,   ${f^{2}(r)=A(r)/(\epsilon \omega^{2})}$ ${\mbox{if } \VV\neq 0}$.}  \\
  \hline
\end{tabular}\vspace{-0.2cm}
\caption{\small{Boundary and junction conditions for Born-Infeld scalar boson stars in spherical coordinates}}
\label{table:schwbcs}
\end{center} 
\end{table}

}


\section{Sample solutions and comments on stability}\label{sec:numer}
The above analysis illustrates the local properties that Born-Infeld scalar condensate must posses in order to have a regular center and proper outer boundary. The Schwarzschild coordinates are locally adequate for the analysis, and were utilized due to their simplicity and geometric interpretation  (the radial coordinate directly measuring the area of 2-spheres). However, for reasons which will become apparent, for a study of the full interior it is better to adopt another coordinate system, which in the case here will be isotropic coordinates. The line element in isotropic coordinates is given by
\begin{align}
{\dd}s^{2} &=-\alpha(\rho)\,{\dd}t^{2} \nonumber \\
 &\quad +\beta(\rho)\left[{\dd}\rho^{2} +\rho^{2}{\dd}\theta^{2}+\rho^{2}\sin^{2}\theta\,{\dd}\phi^{2}\right]. \label{eq:isometric}
\end{align}
If at the center $\beta(\rho)\neq 0$ and finite then the center corresponds to $\rho=0$ from the relation $r=\sqrt{\beta(\rho)}\rho$ and the domain $\rho < 0$ is not relevant for this study. In these coordinates the Schwarzschild metric is furnished by
\begin{align}
 {\dd}s^{2}_{\tinyrmsub{Schw}}&=-\left(\frac{1-M/2\rho}{1+M/2\rho}\right)^{2}{\dd}t^{2} \nonumber \\
 &\quad +\left(1+M/2\rho\right)^{4}\left[{\dd}\rho^{2} +\rho^{2}{\dd}\theta^{2}+\rho^{2}\sin^{2}\theta\,{\dd}\varphi^{2}\right], \label{eq:isoschw}
\end{align}
with $\rho=M/2$ the location of the event horizon. It is worth noting here that the chart (\ref{eq:isoschw}) for $\rho > 0$ actually covers \emph{two} branches outside of the Schwarzschild event horizon. (Mathematically so can the Schwarzschild coordinates, but one would need two coordinate patches.) One branch is $0<\rho< M/2$, where the radius of 2-spheres decreases with increasing $\rho$, and the other is $\rho>M/2$, where the radius of 2-spheres increases with increasing $\rho$ (see the upper-left image in fig.~\ref{fig:bagofgold} for a qualitative depiction). This will turn out to be important for the study.

\begin{figure}[t!]
\begin{center}
\includegraphics[width=1.07\columnwidth]{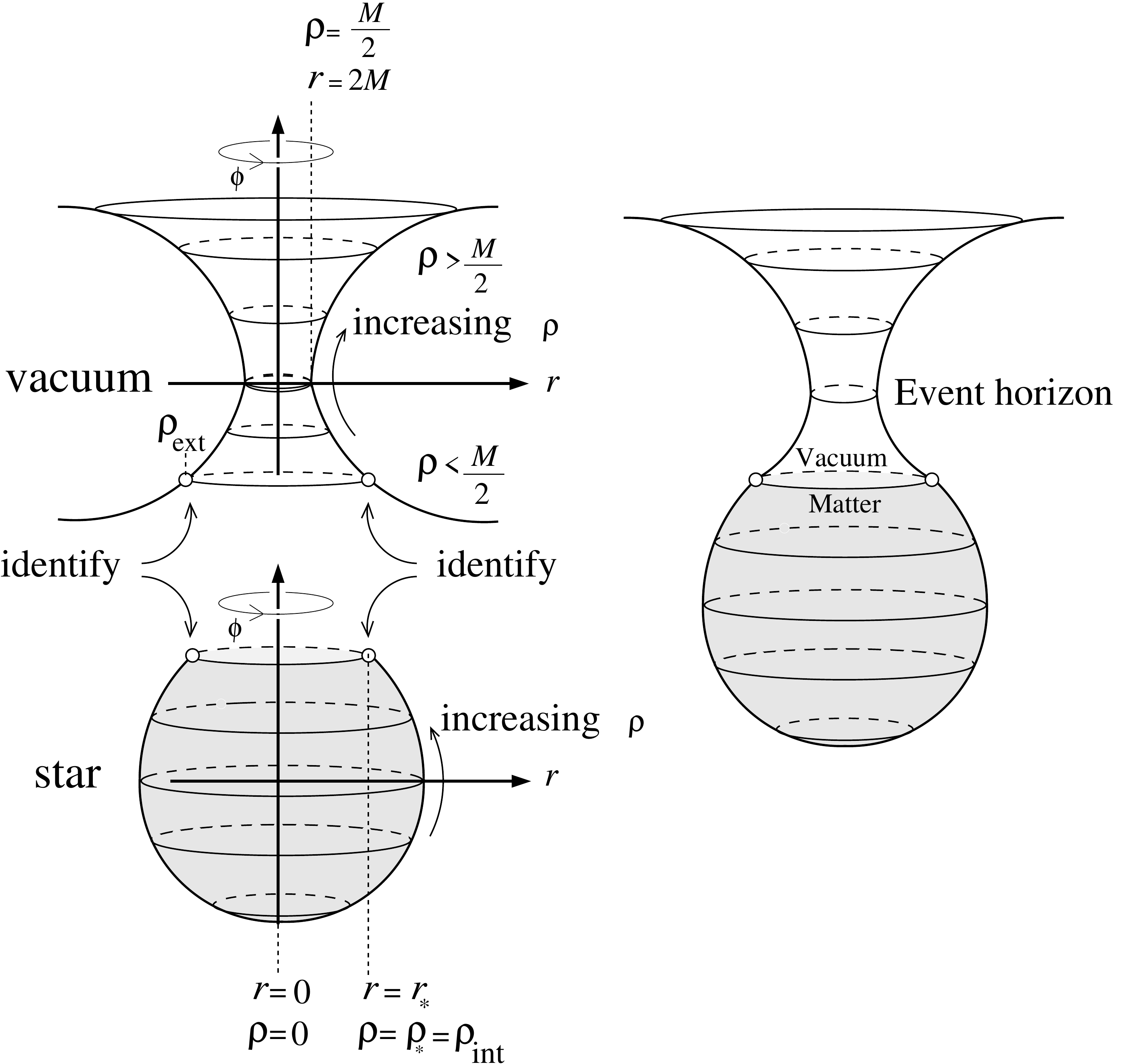}
\end{center}
\vspace*{-5.0mm}\caption{\label{fig:bagofgold} \small{Top-left: A qualitative depiction of the profile in the vacuum spacetime. The isotropic chart Schwarzschild metric (\ref{eq:isoschw}) is capable of covering two vacuum regions, one for $\rho < M/2$ and one for $\rho > M/2$, with a vacuum event horizon at $\rho=M/2$. Bottom left: The matter region. Right: The matter region patched to the vacuum at the junction and a resulting ``bag of gold'' structure. (See text for details.)}}
\end{figure}

\revision{
The left-hand sides of Einstein equations in isotropic coordinates are as follows:
\begin{subequations}\romansubs
\begin{align}
G^{t}_{\;t} & = \frac{\beta''}{\beta^2}
   + \frac{\beta' (8\beta + 3 \rho \beta')}{4 \rho \beta^3}\,,  \label{eq:isoeinsttt}  \\
G^{\rho}_{\;\rho} &  = \frac{\alpha'}{\rho\alpha\beta}
   + \frac{\beta'(2\alpha+\rho\alpha')}{2\rho\alpha\beta^2}
   + \frac{\beta'^2}{4\beta^3}\,, \label{eq:isoeinstrr} \\
G^{\theta}_{\;\theta} & = \frac{1}{4\rho\alpha^2\beta^3} \Big(
     - \rho\beta^2\alpha'^2
     + 2 \alpha \beta^2 ( \alpha' + \rho \alpha'' ) \nonumber \\
     &\quad - 2 \alpha^2 ( \rho \beta'^2 - \beta ( \beta' + \rho \beta'' ))
   \Big) = G^{\phi}_{\;\phi}\,, \label{eq:isoeinstthth}
\end{align} 
\end{subequations}
primes indicating a derivative with respect to $\rho$.
The right-hand sides are functionally identical to (\ref{eq:rho} - \ref{eq:pt}) save for the notational changes $A\to\alpha$, $B\to\beta$ and $r\to\rho$.

The one non-trivial conservation law reads:
\begin{align}
 \nabla_{\mu}T^{\mu}_{\;\rho} & = \frac{1}{2\alpha\beta\rho} \left[\rho  \beta  \alpha^{\prime} \left(T^{\rho}_{\;\rho} -T^{t}_{\;t} \right) +2 \alpha  \left(\rho \beta^{\prime} \left(-T^{\theta}_{\;\theta} +T^{\rho}_{\;\rho} \right)\right.\right. \nonumber \\
 &\qquad +\left.\left. \beta  \left(\rho  T^{\prime \rho}_{\;\rho} +2 T^{\rho}_{\;\rho} -2 T^{\theta}_{\;\theta} \right)\right)\right]\,.   \label{eq:isocons} 
\end{align}
}

Before proceeding we will simply summarize the required junction conditions in isotropic coordinates here without showing their derivations as the methods are similar to those done in the Schwarzschild coordinates. These conditions are listed in table~\ref{table:isobcs}. 
\begin{table}[htbp]
\begin{center}
\begin{tabular}{ |p{3cm}||p{3.5cm}|  }
 \hline
 \multicolumn{2}{|c|}{\makecell{Boundary and junction conditions \\ -isotropic chart-}} \\
 \hline
 Location & Requirement \\
 \hline
 {\vspace{0.4cm}center ($\rho=0$)}   & ${\partial_{\rho}f(\rho)=0}$,\, ${\alpha(\rho)\neq 0}$,\, ${\partial_{\rho}\alpha(\rho)=0}$, \, ${\beta(\rho)\neq 0}$, \,  ${\partial_{\rho}\beta(\rho)=0}$. \\[0.4cm] \hdashline
 {\vspace{1.0cm}matter-vacuum boundary} & {\vspace{0.0cm}${\partial_{\rho}f(\rho)_{|\rho=\rhoint}\neq 0}$ ${\mbox{if } \VV\neq 0}$, \, ${\partial_{\rho}f(\rho)_{|\rho=\rhoint}\not\to\infty}$,\,
 ${[\alpha(\rho)]_{\pm}=0}$, \, ${[\rho^{2}\beta(\rho)]_{\pm}=0}$,  \, ${[\rho\partial_{\rho}\alpha(\rho)]_{\pm}=0}$, \, ${[\rho^{3}\partial_{\rho}\beta(\rho)]_{\pm}=0}$, \, ${f^{2}(\rhoint)=\alpha(\rho)/(\epsilon \omega^{2})}$ ${\mbox{if } \VV\neq 0}$.}  \\
  \hline
\end{tabular}\vspace{-0.2cm}
\caption{\small{Boundary and junction conditions for Born-Infeld scalar boson stars in isotropic coordinates}}
\label{table:isobcs}
\end{center} 
\end{table}

As mentioned above, unlike with Schwarzschild coordinates, in this coordinate chart the value of the coordinate $\rho$ at the matter-vacuum junction does {not} need to be continuous in order to have an acceptable identification. There will therefore generally be two values of $\rho$ at the outer junction, one on the matter-side, which we will refer to below as $\rhoint$, and one on the vacuum side, which we refer to as $\rhoext$.
Knowing the values of the metric and its derivatives in the exterior, as given by metric (\ref{eq:isoschw}), it is a simple matter to determine the values of $M$ (the Schwarzschild mass) and $\rhoext$ from the junction conditions. The values of the metric functions $\alpha(\rhoint)$ and $\beta(\rhoint)$ and $\rhoint$ from the numerical run are given at the material surface by the computational code, and then these values are patched to (\ref{eq:isoschw}) using the junction conditions in table~\ref{table:isobcs}, yielding $M$ and $\rhoext$. Computationally, the material surface is heralded by the vanishing of the radial pressure (\ref{eq:synge}). However, this does not guarantee that the patching is thin-shell free, since one has to then check that the ISLD junction conditions can also be fully met there. All results presented here meet all junction conditions in table~\ref{table:isobcs}.
\revision{We summarize the solution procedure as follows:
\begin{itemize}

\item We utilized standard routines for numerical evolution
of systems of ordinary differential equations
available within the Wolfram Mathematica package.

\item Einstein equations $^t_t$ and $^{\rho}_{\rho}$
and the Born-Infeld field equation are utilized to generate the solutions. As mentioned earlier, the Born-Infeld equation of motion implies that the conservation equation holds. Therefore, with the $^t_t$ and $^{\rho}_{\rho}$, and conservation equation satisfied, the $^{\theta}_{\theta}\;\;(=^{\phi}_{\phi})$ Einstein equation is solved. This is because the conservation law, $\nabla_{\mu}T^{\mu}_{\;\rho}=0$, contains $T^{\theta}_{\;\theta}$ but not its derivatives, and $\nabla_{\mu}G^{\mu}_{\;\rho}\equiv 0 = \nabla_{\mu}T^{\mu}_{\;\rho}$ then enforces that the $^{\theta}_{\theta}$ equation is solved. This has been independently checked for all solutions presented.

The matter domain, where the numerical evolutions took place is from $\rho\approx 0$ (the center of the star) to $\rho=\rhoint$ where the Synge junction condition is met. Then the system is glued properly to the Schwarzschild exterior spacetime (\ref{eq:isoschw}) as described below.

\item Since the equations are singular at $\rho=0$,
the outward evolutions are started from a finite value of $\rho$
which is taken sufficiently close to zero so that the final results are insensitive to its actual value.
This insensitivity was explicitly checked in all cases.

\item At the starting point above we utilize the following initial conditions:
      $$ \alpha = \alpha_0, \quad
         \beta = 1, \quad
         f = f_0, \quad
         \beta' = f' = 0  $$ (the system of equations we chose to solve are first-order in $\alpha$, hence $\alpha'$ does not need to specified).
      Note that the value of $\alpha_0$ is related to the scaling
      of the time coordinate in the interior and can be a posteriori tuned
      to enforce smooth joining at the stellar surface and simultaneously synchronize time with the exterior Schwarzschild time coordinate.

\item As mentioned above, integration is stopped where the Synge boundary condition,
$p_{\rho} = 0$, is met.
At this value of $\rho$, which we refer to as $\rhoint$, the field value, $f$,
is in general found to diverge suddenly to either positive or negative infinity, indicating a breakdown in the evolution.

\item Iteratively, the value of $\omega$ is then adjusted, and the run repeated
in order to find a value which renders finite $f$ at the end-point.
That is, it is possible to fine-tune the value of $\omega$
so that the field value remains finite at $\rho=\rhoint$.
In this sense, the value of $\omega$ can be understood as the eigenvalue,
and the system being solved as the eigenvalue problem.

\item At this stage only the Synge junction condition is guaranteed solved at the junction. The Synge condition is the last one in table \ref{table:isobcs}.  As noted previously though the Synge condition does not guarantee the entirety of the ISLD conditions. From table \ref{table:isobcs} the  remaining junction conditions which must be met for proper gluing are that the quantities
$\alpha(\rho), \, \rho^{2}\beta(\rho),\, \rho\partial_{\rho}\alpha(\rho),\,\rho^{3}\partial_{\rho}\beta(\rho)$ at $\rhoint$ must be equal to their corresponding quantities at $\rhoext$. We have two undetermined parameters, the mass parameter $M$ and the value of $\rhoext$. These are now set by demanding the continuity of these quantities. Although we have four conditions and only two parameters, the four conditions actually boil down to two independent conditions, due to the fact that the Synge junction conditions have already been satisfied. In other words, the mass parameter, $M$, and $\rhoext$ are taken to be what they \emph{must} be in order to have a proper gluing of the interior spacetime, terminating at $\rho=\rhoint$, to the exterior Schwarzschild geometry at $\rho=\rhoext$. We note here that the mass parameter set in this way actually agrees with the integral of the energy density taken over the matter region as in (\ref{eq:schwmass}). Therefore, with a proper gluing, everything is consistent.

\item Once the above conditions have been met the interior spacetime is properly glued to the exterior Schwarzschild geometry, where automatically all components of the stress-energy tensor are zero. The potential $\VV$ in the vacuum region is therefore zero by construction. Therefore, if at $\rhoint$ the potential does not vanish (as discussed previously it is not required to) then the aforementioned jump discontinuity in $\VV$ is present here.
\end{itemize}

}

Now, if the junction of the star occurs where $\rhoext < M/2$, it is patching to vacuum in the lower funnel of the upper-left of fig.~\ref{fig:bagofgold}, where two-spheres are decreasing as $\rho$ increases. If the junction occurs where $\rhoext > M/2$ the patching is occurring in the upper funnel of fig.~\ref{fig:bagofgold}, where two-spheres are increasing as $\rho$ increases. (See the Appendix for some details on this.)  If we have the first scenario ($\rhoext<M/2$), the resulting geometry is that of a ``bag of gold'' spacetime \cite{ref:wheelbagofgold}, \cite{ref:bagofgold2}, and we see from the figure that in that case the boson star is hidden behind an event horizon. This event horizon is locally equivalent to the Schwarzschild wormhole \cite{ref:flamm_parabola} which is known to be unstable  \cite{ref:wheelerbook}, \cite{ref:kruskal}. Therefore it is likely that this horizon throat will collapse due to the same instability present in the Schwarzschild wormhole, sealing off the Born-Infeld star from the exterior.

\subsection{\boldmath{$\VV=\mu^{2}\phi^{*}\phi$}}
Many different potentials $\VV$ were utilized but we present here only two for brevity. The first potential we present is the analog of the mass term in the canonical scalar field. That is, the potential to be used will have the form $\VV=\mu^{2}\phi^{*}\phi$. Of course, for the Lagrangian density (\ref{eq:tachylag}) this will not be a true mass term, but it is the simplest non-trivial potential that can be used to illustrate the properties of these objects. Many numerical runs with this potential were performed with various initial conditions and parameters and all of them qualitatively resembled the graphs that will be shown here.

Fig.~\ref{fig:massfig1} presents various quantities related the Schwarzschild mass for the potential $\VV=\mu^{2}\phi^{*}\phi$ (see figure caption for description). Each evolution starts by choosing a specific value of the field function $f(\rho)$ at the center while meeting the central boundary conditions in table~\ref{table:isobcs}. Then the parameter $\omega$ is tuned until a scenario is found which is evolved to a value of $\rho$ where $p_{r}$ vanishes, and then the remaining junction conditions are checked to be satisfied. The figure illustrates 128 different scenarios. Since the Born-Infeld action (\ref{eq:tachylag}) is globally U(1) invariant, the negative $f(0)$ sector is a mirror image of the positive sector and is therefore not shown.

\begin{figure}[htbp]
\begin{center}
\includegraphics[width=0.95\columnwidth]{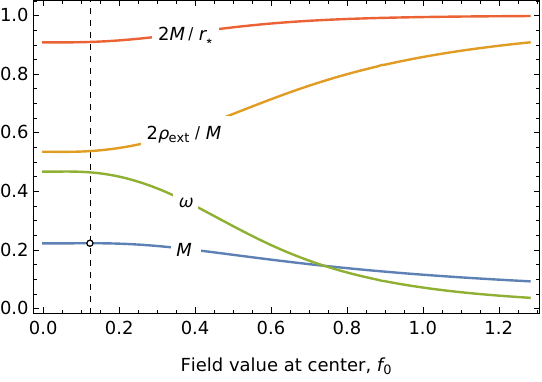}
\end{center}
\vspace*{-5.5mm}\caption{\label{fig:massfig1} {\small{Various evolutions of the  ``mass'' potential, $\VV=\mu^{2}\phi^{*}\phi$. The horizontal axis represents $f(0)$, the field value at the center. The plots are: Surface compactness (red), $2\rhoext/M$, (orange) which is always smaller than 1 so the patching is occurring at a surface where ${\dd}r/{\dd}\rho<0$, the frequency (green) and the Schwarzschild mass of the star (blue). The small circle represents the maximum mass run. See main text for details.}}}
\end{figure}

Of particular interest in the plots
are the graphs of the surface compactness $2M/r$ and the ratio $2\rhoext/M$.
By increasing the central field value one can generate solutions
whose surface compactness approach arbitrarily close to $1$,
and hence there is no Buchdahl bound on these stars. Also, $2\rhoext/M$ is always less than one, indicating that $\rhoext<M/2$ and therefore the patching to vacuum is always occurring in the lower branch of the Flamm's parabola in fig.~\ref{fig:bagofgold}. As mentioned previously, with all the potentials studied, all with many different values of $f(0)$, there was never a run which patched at $\rhoext>M/2$. This means that all runs have the bag of gold structure, and are hidden behind the event horizon. Analytically we could find no argument for why $\rhoext>M/2$ is not allowed. This analysis is shown in the Appendix.

In fig.~\ref{fig:massfig2} we illustrate the stellar mass $M$
to surface radius $r_*$ relationship for the Born-Infeld stars
with potential $\VV=\mu^{2}\phi^{*}\phi$
for different values of the central field $f(0)=f_{0}$.
In the realm of astrophysics these plots are interesting
as they may reveal the onset of the dynamical instability
of the stellar configurations with respect to radial perturbations
\cite{ref:wheelerstabilitybook}.
Usually, e.g.\ for relativistic polytropic spheres
and also for some anisotropic stars \cite{Horvat:2010xf},
the $M \to 0$ branch is stable,
while the first maximum of the mass indicates the onset of instability,
sometimes referred to as the relativistic instability.
However, in the context of Born-Infeld stars
we have unusual features regarding the definition of mass, which we discuss below. In the figure it \emph{may} be that the left part of the line indicates a stable branch and the right part an unstable branch. However, our analysis at the present stage does now allow us to draw
firm conclusions regarding the stability of these objects.

\begin{figure}[htbp]
\begin{center}
\includegraphics[width=0.95\columnwidth]{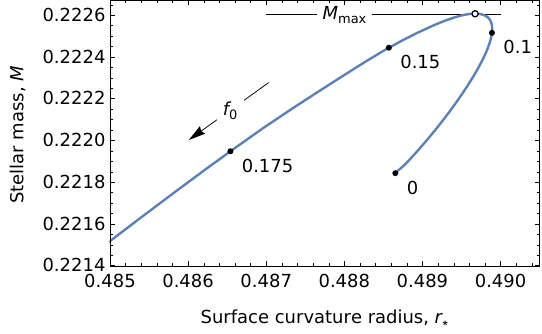}
\end{center}
\vspace*{-5.5mm}\caption{\label{fig:massfig2} {\small{The Schwarzschild mass-to-radius curve for Born-Infeld stars with the potential $\VV=\mu^{2}\phi^{*}\phi$. The numbers along the plot indicate the central field value $f_{0}=f(0)$. See main text for details.}}}
\end{figure}

In Figs \ref{fig:massfig3} and \ref{fig:massfig4b}
we plot one specific member of the family of solutions studied above.
We chose the solution giving the maximal mass $M=M_{\max}\simeq0.223$
which is indicated by small circles
in Figs \ref{fig:massfig1} and \ref{fig:massfig2},
although qualitatively all members of the $\VV=\mu^{2}\phi^{*}\phi$, $\mu=1$,
family have a resemblance to the given plots.
The parameter values for this solution are
$f_0 \simeq 0.123$, $\omega \simeq 0.464$, $2M/r_* \simeq 0.909$.
In fig.~\ref{fig:massfig3} we illustrate quantities related to the metric:
   $\alpha(\rho)$, $\beta(\rho)$,
   $1/B(r(\rho))$,
   the mass function $m(r(\rho))$,
   along with the local compactness within the star $2m(r(\rho))/r(\rho)$,
   and the 2-spheres radius $r(\rho)=\rho\sqrt{\beta(\rho)}$.
The vertical dashed line indicates where $B(r(\rho))$ ($g_{rr}$ of the Schwarzschild coordinates) becomes infinite. This is where the radii of 2-spheres transitions from increasing to decreasing. Although it is not easy to see, the green $r(\rho)$ line in the plot is decreasing to the right of this vertical dashed line. This point is \emph{not} an event horizon since for an event horizon in these coordinate systems we require $\alpha(\rho)=0$ (which is exactly the same condition as $A(r)=0$ in Schwarzschild coordinates.)

\begin{figure}[htbp]
\begin{center}
\includegraphics[width=0.95\columnwidth]{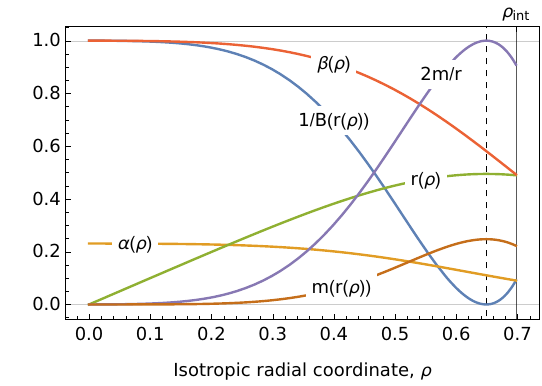}
\end{center}
\vspace*{-5.5mm}\caption{\label{fig:massfig3}
{\small{Quantities relevant to the compact object
with $\VV=\mu^{2}\phi^{*}\phi$, with $\mu=1$ and $f_{0}=0.123$.
Displayed are the metric functions $\alpha(\rho)$ and $\beta(\rho)$,
along with the inverse of the Schwarzschild metric function, $1/B(r(\rho))$,
the local compactness, mass function, and 2-sphere radius.
See main text for details.}}}
\end{figure}

In fig.~\ref{fig:massfig4b} we show the components of the stress-energy tensor (\ref{eq:rho}-\ref{eq:pt}) (but the quantities computed from the isotropic coordinate chart) along with the field value $f(\rho)$.

\begin{figure}[h!]
\begin{center}
\includegraphics[width=0.95\columnwidth]{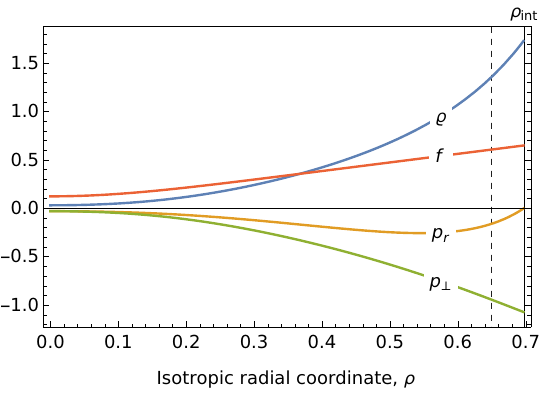}
\end{center}
\vspace*{-5.5mm}\caption{\label{fig:massfig4b} {\small{Stress-energy components and field value for the potential $\VV=\mu^{2}\phi^{*}\phi$,
with $\mu=1$ and $f_{0}=0.123$.}}}
\end{figure}

\revision{
Finally, in order to show explicitly that the ISLD junction conditions are met, we can convert our metric expressions from isotropic coordinates back into the spherical coordinates of section \ref{sec:junct}, and show that the appropriate continuities from table \ref{table:schwbcs} are met. This is illustrated in fig.~\ref{fig:massfig5}.
\begin{figure}[h!]
\begin{center}
\includegraphics[width=0.95\columnwidth]{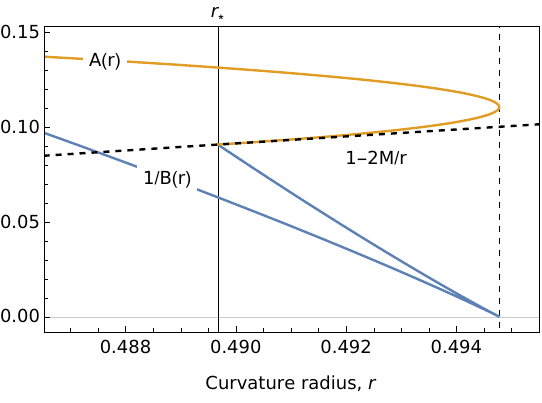}
\end{center}
\vspace*{-5.5mm}\caption{\label{fig:massfig5} {\small{Continuity of $A(r)$, $B(r)$, and $\partial_{r}A(r)$ and the stellar boundary for the potential $\VV=\mu^{2}\phi^{*}\phi$ with $\mu=1$ and $f_{0}=0.123$. Note that $A(r)$ and $1/B(r)$ are joining the function $1-2M/r$ (dashed) which is also shown in the figure. Also note that the ISLD conditions do \emph{not} demand continuity of $\partial_{r}B(r)$.}}}
\end{figure}
}

Before proceeding we should caution on the meaning of the mass parameter $M$ here. The mass $M$ is the quantity which is present in the Schwarzschild metric
\begin{equation}
 {\dd}s^{2}=-\left(1-\frac{2M}{r}\right){\dd}t^{2} + \frac{{\dd}r^{2}}{1-\frac{2M}{r}} + r^{2}\,{\dd}\theta^{2} +r^{2}\sin^{2}\theta\, {\dd}\varphi^{2}\,. \label{eq:schwmet}
\end{equation}
It is also the same $M$ which occurs in metric (\ref{eq:isometric}).
In single-valued scenarios,
i.e.~if $r$ is a monotonically increasing function of $\rho$,
this $M$ is explicitly calculated in the Schwarzschild coordinate chart via
\begin{equation}
M:=4\pi\int_{0}^{r_{*}} \varrho(r^{\prime})\,r^{{\prime}2}\, {\dd}r^{\prime}\,, \label{eq:schwmass}
\end{equation}
and yields the ADM mass of the spacetime. However, when $\varrho(r^{\prime})$ is multivalued in this coordinate chart, as it is with the bag of gold spacetime, the interpretation of $M$ as an integral of the energy density over the (coordinate) volume of the star becomes ambiguous. Birkhoff's theorem in vacuum still holds (its most well-known proof is local) and therefore the vacuum metric is still (\ref{eq:isoschw}) and (\ref{eq:schwmet}). Now, when we have the junction conditions of table~\ref{table:isobcs} and apply them at the location where $p_{r}=0$, they yield a \emph{unique} value for the constant $M$. The question then is, what is this $M$ in relation to the integral (\ref{eq:schwmass})? It turns out, perhaps surprisingly, that in the bag of gold spacetime the integral which gives the $M$ that coincides with the Schwarzschild metric $M$ turns out to be
\begin{align}
 M& =4\pi\underset{D_{1}}{\int_{0}}^{\rmax} \varrho(r^{\prime})\,r^{{\prime}2}\, {\dd}r^{\prime} + 4\pi\underset{D_{2}}{\int_{\rmax}}^{r_{*}} \varrho(r^{\prime})\,r^{{\prime}2}\, {\dd}r^{\prime}  \label{eq:massints} \\[-0.3cm]
 & = 4\pi\underset{D_{1}}{\int_{0}}^{\rmax} \varrho(r^{\prime})\,r^{{\prime}2}\, {\dd}r^{\prime} - 4\pi\underset{D_{2}}{\int_{r_{*}}}^{\rmax} \varrho(r^{\prime})\,r^{{\prime}2}\, {\dd}r^{\prime}\,. \nonumber 
\end{align}
Here the $D_{1/2}$ notation indicates the two different regions of the bag of gold (the branch where $r$ is increasing with increasing $\rho$ and the branch where $r$ is decreasing with increasing $\rho$), and $\rmax$ is the radius of 2-spheres where the bag of gold is the widest in fig.~\ref{fig:bagofgold}, which is the border between these two regions. Since $\rmax > r_{*}$ for the bag of gold, the second term decreases the value of $M$ if the energy density is positive. Therefore, in this type of geometry the Schwarzschild mass is not a good measure of the total mass making up the star as it is not an additive sum of the total energy density within the star. This also explains why in fig.~\ref{fig:massfig3} $m(r(\rho))$ is decreasing as $\rho$ increases beyond the point corresponding to $\rmax$, even though the energy density is positive there. As an extra consistency check, it has been verified that $m(\rhoint)$ is equal to the constant $M$ in (\ref{eq:isoschw}) determined from the junction conditions.

\subsection{\boldmath{$\VV=a_{0}e^{-\kappa\sqrt{\phi^{*}\phi}}$}}
The second potential we present is of the form
\begin{equation}
\VV=a_{0}e^{-\kappa\sqrt{\phi^{*}\phi}}\,, \label{eq:senpot}
\end{equation}
for $a_{0}$ and $\kappa$ positive constants. This particular potential is motivated by some of the original string theory work where the tachyonic scalar field was presented \cite{ref:senBIjhep}, \cite{ref:sentachy}, \cite{ref:senBImpla}. In the theory of
D-branes with real scalar field it is desirable that the potential has its minimum at $\phi\rightarrow \pm \infty$ and its maximum at $\phi=0$. The often-used form of the potential for $\phi \geq 0$ in D-brane studies is $V(\phi)=e^{-\sqrt{2}\phi}$ and (\ref{eq:senpot}) is the simplest extension of this potential to the complex regime. We caution that there could be mathematical issues as $\phi$ and $\phi^{*}$ go to zero with such a potential but we will see here that the evolution can proceed without major impediment. The presentation here mimics that of the ``mass'' potential and therefore we do not give all the details.

In figs.~\ref{fig:senfig1}-\ref{fig:senfig5} we display
quantities analogous to figs.~\ref{fig:massfig1}-\ref{fig:massfig5}.
The general qualitative features are the same,
and turn out to be the same for all potentials studied,
which is why we only present two types of potentials here.
One contrast between this potential and the mass potential
is in the mass versus radius plot
of figs.~\ref{fig:massfig2} and \ref{fig:senfig2}.
With the potential (\ref{eq:senpot})
we do not find the maximum of the mass $M$.
Therefore, according to the usual interpretation of the mass-to-radius curve,
all solutions obtained here are either stable or unstable.
At the present stage we can not tell which of the two is the case.

\begin{figure}[htbp]
\begin{center}
\includegraphics[width=0.95\columnwidth]{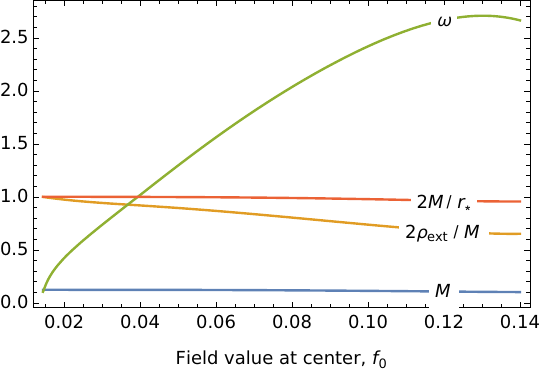}
\end{center}
\vspace*{-5.5mm}\caption{\label{fig:senfig1} {\small{Various compact solutions for the potential $\VV=a_{0}e^{-\kappa\sqrt{\phi^{*}\phi}}$ for $a_{0}=1$ and $\kappa=\sqrt{2}$ as a function of central field value $f(0)$.
Depicted: Surface compactness (red),
$2\rhoext/M$ (orange) which is always smaller than 1
so the patching is occurring a surface where ${\dd}r/{\dd}\rho<0$,
the frequency (green) and the Schwarzschild mass of the star (blue).
The masses of all compact objects here are almost the same.}}}
\end{figure}

\begin{figure}[h!]
\begin{center}
\includegraphics[width=0.95\columnwidth]{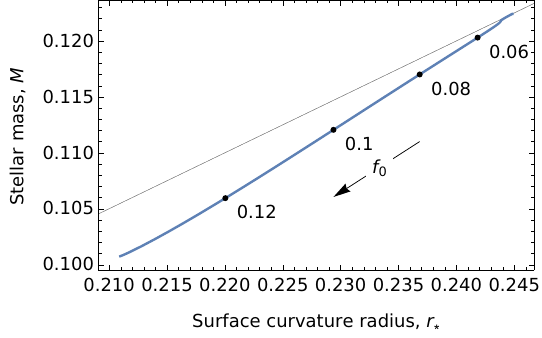}
\end{center}
\vspace*{-5.5mm}\caption{\label{fig:senfig2} {\small{The Schwarzschild mass-to-radius curve for Born-Infeld stars with the potential $\VV=a_{0}e^{-\kappa\sqrt{\phi^{*}\phi}}$ for $a_{0}=1$ and $\kappa=\sqrt{2}$. The thin gray line represents $r=2M$. Note that the solutions with larger radii approach arbitrarily close to $r=2M$.}}}
\end{figure}

\begin{figure}[h!]
\begin{center}
\includegraphics[width=0.95\columnwidth]{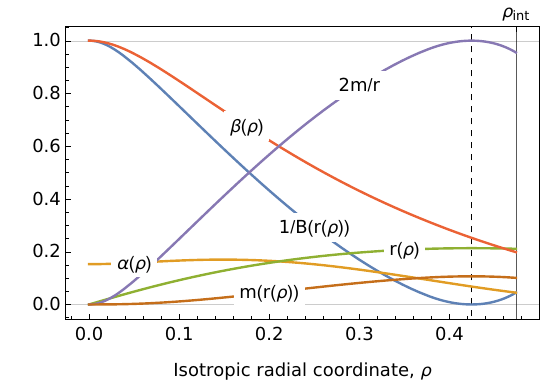}
\end{center}
\vspace*{-5.5mm}\caption{\label{fig:senfig3} {\small{Quantities relevant to the compact object with $\VV=a_{0}e^{-\kappa\sqrt{\phi^{*}\phi}}$ for $a_{0}=1$ and $\kappa=\sqrt{2}$ and $f_{0}=0.14$.
Displayed are the metric functions $\alpha(\rho)$ and $\beta(\rho)$,
along with the inverse of the Schwarzschild metric function, $1/B(r(\rho))$,
the local compactness, mass function, and 2-sphere radius.}}}
\end{figure}

\begin{figure}[!h]
\begin{center}
\includegraphics[width=0.95\columnwidth]{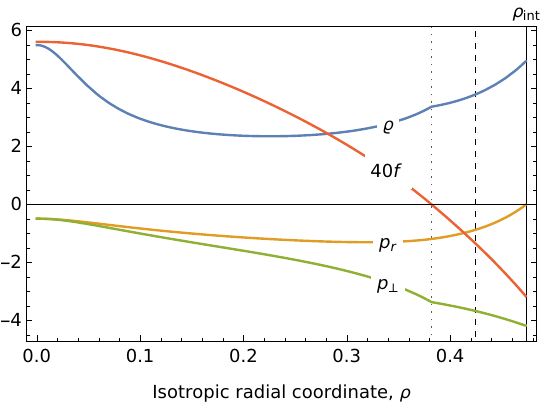}
\end{center}
\vspace*{-5.5mm}\caption{\label{fig:senfig4b} {\small{
Stress-energy components and field value
(rescaled by a factor of 40 for clarity)
for the potential  $\VV=a_{0}e^{-\kappa\sqrt{\phi^{*}\phi}}$
with $a_{0}=1$ and $\kappa=\sqrt{2}$ and $f_{0}=0.14$.
There is a small kink in some quantities, indicated by the dotted line,
where the field is zero
due to the presence of the function $\sqrt{\phi^{*}\phi}$ in the potential.}}}
\end{figure}

\revision{The satisfaction of the ISLD junction conditions from table \ref{table:schwbcs} are met and this is is illustrated in fig.~\ref{fig:senfig5}.
\begin{figure}[h!]
\begin{center}
\includegraphics[width=0.95\columnwidth]{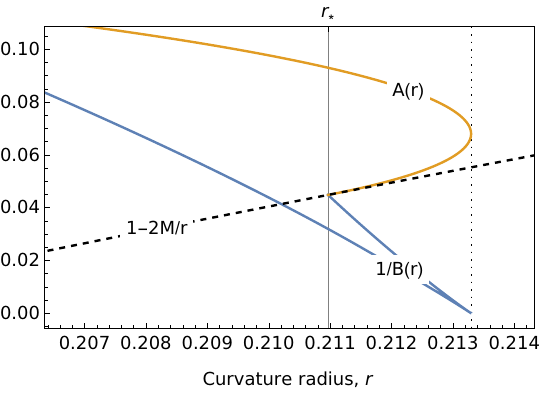}
\end{center}
\vspace*{-5.5mm}\caption{\label{fig:senfig5} {\small{Continuity of $A(r)$, $B(r)$, and $\partial_{r}A(r)$ and the stellar boundary for the potential $\VV=a_{0}e^{-\kappa\sqrt{\phi^{*}\phi}}$ with  $a_{0}=1$ and $\kappa=\sqrt{2}$ and $f_{0}=0.14$.}}}
\end{figure}
}


\section{Some comments on the cosmological constant} \label{sec:cosconst}
We briefly comment here on how a possible cosmological constant could affect the solutions. In the presence of a cosmological constant the vacuum to be patched to will no longer be the Schwarzschild vacuum but instead the Schwarzschild-(anti) deSitter or Kottler vacuum. The Synge condition, requiring continuity of the pressure at the matter-vacuum junction, would mean that the net radial pressure (material pressure plus cosmological term) should be continuous at the boundary. This fact could possibly alleviate the ${\dd}r/{\dd}\rho <0$ condition where the patching occurs. One reason is that the shape of the vacuum profile curve will no longer be the Flamm's parabola of fig.~\ref{fig:bagofgold} and it may be possible to patch where ${\dd}r/{\dd}\rho \not{<} 0$. Also, if one allows for the cosmological term to differ inside and outside of the stellar boundary then the field's radial pressure would not have to vanish in order to have continuity of the net radial pressure. This would change the location of the vacuum-matter boundary and it may be possible that the patching could occur where ${\dd}r/{\dd}\rho > 0$. A proper study would need to be performed to see if such scenarios can actually be realized.


\section{Concluding remarks}\label{sec:conc}

In this manuscript a detailed study of Born-Infeld scalar compact objects (``Boson stars''), as motivated by Born-Infeld theory and string theory, was presented. The structure of these condensates turns out to be surprisingly rich. First, unlike their canonical scalar field counter-parts, these stars can be truly compact, meaning of finite extent, for arbitrary potentials. The boundary and junction conditions at the center of the star, as well as at the matter-vacuum junction were studied. It turns out that there are several ways that in principle the matter-vacuum junction conditions of Israel-Sen-Lanczos-Darmois and those of Synge-O'Brien can be met. There are two natural classes of solutions that arise analytically: One class is where the surface of the compact object is in a region where the radii of 2-spheres is increasing as one moves outward from the star, and the other is in a region where the 2-sphere radii are decreasing as one moves outward, yielding a bag of gold spacetime.

In practice, meaning here using numerics, the only successful runs meeting all junction conditions were for the latter scenario; the bag of gold. Also, the junction conditions were always met under the condition that the potential did not vanish at the boundary. We further analyzed this in the attempt to determine whether the junctions conditions force one scenario over the other,  but the results were inconclusive.  That is, there was no mathematical reason that could be found as to why the bag of gold structure was preferred over the usual geometry generated by stars. These calculations are presented in the appendix. Parts of the strong-energy condition, viz. non-negativity of $\varrho+p_{r/\perp}$ and/or $\varrho+p_{r}+2p_{\perp}$, are violated near the surface, and this is surely a source of the peculiar geometry, but no reason could be found as to why this violation must occur. Since most of the analysis presented here is local, it is likely that, given a set of acceptable boundary conditions at the center, the evolution equations governing the metric in the bulk of the star force a contracting 2-sphere geometry in the vicinity of the outer junction. This situation would not easily be detected by the local analysis of the surface junction conditions performed here.

For the bag of gold structure, the star is hidden behind a vacuum event horizon. It is conjectured that, being locally equivalent to the Schwarzschild wormhole, this horizon throat would be unstable, quickly cutting off the star from the exterior universe.

Finally, some comments were made regarding the addition a cosmological constant to the scenario, and how it may affect the matter-vacuum patching and the shrinking 2-sphere condition.

\vspace{0.7cm}
\section*{Acknowledgments}
This work is partially supported by the VIF program of the University of Zagreb. A.D. is grateful to the University of Zagreb, Department of Applied Physics at FER for kind hospitality during the production of this work. \revision{We also thank the anonymous referee for comments and suggestions which have greatly clarified the manuscript.}

\PRLsep
\vspace{-0.080cm}

{\small
\bibliography{Born_Infeld_compact_objects}}


\appendix
\setcounter{equation}{0}
\renewcommand\theequation{A.\arabic{equation}}

\section*{Appendix - $\bm{{\dd}r/{\dd}\rho}$ at the matter-vacuum boundary}
As mentioned in the main text, the computational models always patch to the matter-vacuum junction in the region where ${\dd}r/{\dd}\rho<0$. However, the analysis indicates no preferred sign for ${\dd}r/{\dd}\rho$ at the junction. In this appendix we show the calculations illustrating this.

The junction conditions of table \ref{table:isobcs} imply that at the junction surface, if $\VV\neq 0$ we must have
\begin{equation}
 \alpha(\rhoint) =\omega^{2}f(\rhoint)^2\,. \label{eq:appalpha}
\end{equation}
We also have the relationship between the $r$ coordinate and the $\rho$ coordinate
\begin{equation}
 r(\rho)=\sqrt{\beta(\rho)}\,\rho\,, \label{eq:appr}
\end{equation}
which gives for ${\dd}r/{\dd}\rho$:
\begin{equation}
 \frac{{\dd}r}{{\dd}\rho}=\frac{\rho}{2\sqrt{\beta(\rho)}} \frac{{\dd}\beta(\rho)}{{\dd}\rho} +\sqrt{\beta(\rho)}\,. \label{eq:appdrdrho}
\end{equation}
Multiplying the above by $\rho^{2}\sqrt{\beta(\rho)}$ yields
\begin{equation}
\rho^{2}\sqrt{\beta(\rho)} \frac{{\dd}r}{{\dd}\rho}=\frac{\rho^{3}}{2} \frac{{\dd}\beta(\rho)}{{\dd}\rho} +\rho^{2}\beta(\rho)\,. \label{eq:appdrdrho2}
\end{equation}
Now, the above calculation is to be performed \emph{inside} the star, but note that the ISLD junction conditions tell us that, at the junction,
\begin{equation}
\begin{split}
 \beta(\rhoint)\rhoint^{2} & = \beta(\rhoext)\rhoext^{2}\, \quad \mbox{and, } \\[0.2cm]
 \rhoint^{3}\frac{{\dd}\beta(\rho)}{{\dd}\rho}_{|\rho=\rhoint}  & =  \rhoext^{3}\frac{{\dd}\beta(\rho)}{{\dd}\rho}_{|\rho=\rhoext}\,,
\end{split}
\end{equation}
so, \emph{at the junction} (only), we can write (\ref{eq:appdrdrho2}) as
\begin{equation}
\underbrace{\rho^{2}\sqrt{\beta(\rho)}}_{> 0} \frac{{\dd}r}{{\dd}\rho}=\frac{\rhoext^{3}}{2} \frac{{\dd}\beta(\rho)}{{\dd}\rho}_{|\rho=\rhoext}\!\! +\rhoext^{2}\beta(\rhoext)\,, \label{eq:appdrdrho3}
\end{equation}
which, using the Schwarzschild values (\ref{eq:isoschw}) for the external quantities on the right-hand side and simplifying yields
\begin{equation}
\underbrace{\rho^{2}\sqrt{\beta(\rho)}}_{> 0} \frac{{\dd}r}{{\dd}\rho}=\frac{(2\rhoext+M)^{3}(2\rhoext-M)}{16\rhoext^2}\,. \label{eq:appdrdrho4}
\end{equation}
Therefore, the sign of ${\dd}r/{\dd}\rho$ at the junction depends on whether the right-hand side of (\ref{eq:appdrdrho4}) is positive or negative. This criterion proves to be inconclusive however. We note that the above only tells us what we already know from fig.~\ref{fig:bagofgold}, namely that if $\rhoext<M/2$ then ${\dd}r/{\dd}\rho$ there is negative and if $\rhoext>M/2$ then ${\dd}r/{\dd}\rho$ there is positive.

Now, another of the ISLD junction conditions from table~\ref{table:isobcs} is the continuity of $\alpha(\rho)$ at the junction. Therefore, from (\ref{eq:appalpha}) and (\ref{eq:isoschw}) we get
\begin{equation}
\omega^{2}f^{2}(\rhoint)  = \left({\frac{1-\frac{M}{2\rhoext}}{1+\frac{M}{2\rhoext}}}\right)^{2}\,. \label{eq:appcontalph}
\end{equation}
Solving the above for the mass gives two possible masses:
\begin{equation}\label{eq:appmasses}
\begin{split}
 M_{1} & =\frac{2\rhoext(\omega f(\rhoint)+1)}{1-\omega f(\rhoint)}\, \quad \mbox{and,}\\[0.2cm]
 M_{2} & =\frac{2\rhoext(1- \omega f(\rhoint))}{1+\omega f(\rhoint)} \,,
\end{split}
 \end{equation}
which, due to positivity of the mass yields the restriction
\begin{equation}
 -1 < \omega f(\rhoint) < +1\,. \label{eq:appomegafrestrict}
\end{equation}

However, putting both of these masses into the right-hand side of (\ref{eq:appdrdrho4}) gives the following right-hand sides
\begin{equation}
 \frac{-16 \rhoext^{2}f(\rhoint)\omega}{(\omega f(\rhoint) -1)^{4}}\,, \quad \frac{16 \rhoext^{2}f(\rhoint)\omega}{(\omega f(\rhoint) +1)^{4}} \label{eq:appmassdrdrho}
\end{equation}
which again is inconclusive in forcing the sign of ${\dd}r/{\dd}\rho$ since both can be positive or negative in the range (\ref{eq:appomegafrestrict}).

Finally, although redundant, we can solve (\ref{eq:appcontalph}) for $\rhoext$ to see if $\rhoext$ is greater or less than $M/2$. Doing this yields
\begin{equation}\label{eq:apprhos}
\begin{split}
\rhoext & =\frac{M}{2}\, \frac{(1-\omega f(\rhoint))}{(\omega f(\rhoint)+1)}\, \quad \mbox{and,}\\[0.2cm]
\rhoext & =\frac{M}{2}\, \frac{(1+\omega f(\rhoint))}{(1-\omega f(\rhoint))} \,,
\end{split}
\end{equation}
which is also inconclusive since in the range (\ref{eq:appomegafrestrict}) these may be both greater or less than $M/2$.

One more attempt at figuring out whether 2-spheres must be shrinking as $\rho$ increases near, but in the interior of, the junction is by studying the Raychaudhuri equation for the expansion scalar, $\Theta$, which for vanishing vorticity is given by \cite{ref:raychaud1}, \cite{ref:raychaud2}
\begin{equation}
 \frac{\dd\Theta(\tau)}{\dd \tau}=-\frac{1}{3}\Theta^{2}(\tau)-\sigma^{\mu\nu}\sigma_{\mu\nu} +\nabla_{\mu}\left(u^{\nu}\nabla_{\nu}u^{\mu}\right) - R_{\mu\nu}u^{\mu}u^{\nu}\,. \label{eq:apraychaud1}
\end{equation}
Here $\sigma_{\mu\nu}$ is the shear tensor with the property that $\sigma^{\mu\nu}\sigma_{\mu\nu}\geq 0$ and $u^{\mu}$ is the 4-velocity of a congruence of particle worldlines. Let us consider geodesics, so that the $\nabla_{\mu}\left(u^{\nu}\nabla_{\nu}u^{\mu}\right)$ is zero. Also, assuming we neglect $\Theta(s)=0$, we will multiply the equation by $\Theta^{-2}(\tau)$ yielding
\begin{equation}
 \frac{\dd\Theta^{-1}(\tau)}{\dd \tau}=\frac{1}{3}+\underbrace{\left[\sigma^{\mu\nu}\sigma_{\mu\nu} + R_{\mu\nu}u^{\mu}u^{\nu}\right]}_{=:f(\tau)} \Theta^{-2}(\tau)\,. \label{eq:apraychaud2}
\end{equation}
Integrating the above equation yields
\begin{equation}
 \Theta^{-1}(\tau)=\Theta^{-1}(\tau_{0}) + \frac{1}{3}\left(\tau-\tau_{0}\right) +\int_{\tau_{0}}^{\tau} f(\tau^{\prime}) \Theta^{-2}(\tau^{\prime})\, \dd \tau^{\prime}\,. \label{eq:apraysol}
\end{equation}
In principle one could glean from this expression if we have expansion or contraction of geodesic congruence for particles with 4-velocities moving in the positive $\rho$ direction inside the star near the matter-vacuum junction, provided $\tau$ is not too large. If this solution focuses (vs defocusing) for increasing $\rho$ then the worldlines are attempting to converge as $\rho$ increases, reminiscent of what happens if the patching is at the bottom of the Flamm parabola in fig.~\ref{fig:bagofgold}. If it is defocusing for increasing $\rho$, then the situation is reminiscent of a patching on the upper-side of the Flamm parabola. However, the sign of the last term is not determined in general since $f(\tau^{\prime})$ contains the Ricci tensor. Let us write the Ricci tensor, using the Einstein equations, as
\begin{equation}
 R_{\mu\nu}=8\pi\left[T_{\mu\nu}-\frac{1}{2}T^{\alpha}_{\;\alpha} g_{\mu\nu}\right]\,. \label{eq:apriccistress}
\end{equation}
However, we notice from (\ref{eq:rho}) - (\ref{eq:pt}) (even in the isotropic chart) that all components of the right-hand side of (\ref{eq:apriccistress}) are proportional to $\VV$, and for different potentials there is no guarantee that the potential has one sign or the other near the matter-vacuum junction, and consequently  nor can we say anything in general about the sign (and size) of the Ricci tensor term in (\ref{eq:apraysol}). Therefore, we cannot in general state that we must have convergence or divergence near the matter-vacuum junction.




\end{document}